\documentclass[12pt]{article}
\usepackage{amssymb}
\renewcommand{\theequation}{\arabic{section}.\arabic{equation}}

\begin{document}

%************************** Text Begins here ******************************

%  Greek letters

\def\a{\alpha}
\def\b{\beta}
\def\d{\delta}
\def\e{\epsilon}
\def\g{\gamma}
\def\h{\mathfrak{h}}
\def\k{\kappa}
\def\l{\lambda}
\def\o{\omega}
\def\p{\wp}
\def\r{\rho}
\def\t{\tau}
\def\s{\sigma}
\def\z{\zeta}
\def\x{\xi}
\def\V={{{\bf\rm{V}}}}
 \def\A{{\cal{A}}}
 \def\B{{\cal{B}}}
 \def\C{{\cal{C}}}
 \def\D{{\cal{D}}}
\def\G{\Gamma}
\def\K{{\cal{K}}}
\def\O{\Omega}
\def\R{\bar{R}}
\def\T{{\cal{T}}}
\def\L{\Lambda}
\def\f{E_{\tau,\eta}(sl_2)}
\def\E{E_{\tau,\eta}(sl_n)}
\def\Zb{\mathbb{Z}}
\def\Cb{\mathbb{C}}

\def\R{\overline{R}}
% Shorthands for \begin{equation} and the like

\def\beq{\begin{equation}}
\def\eeq{\end{equation}}
\def\bea{\begin{eqnarray}}
\def\eea{\end{eqnarray}}
\def\ba{\begin{array}}
\def\ea{\end{array}}
\def\no{\nonumber}
\def\le{\langle}
\def\re{\rangle}
\def\lt{\left}
\def\rt{\right}

\newtheorem{Theorem}{Theorem}
\newtheorem{Definition}{Definition}
\newtheorem{Proposition}{Proposition}
\newtheorem{Lemma}{Lemma}
\newtheorem{Corollary}{Corollary}
\newcommand{\proof}[1]{{\bf Proof. }
        #1\begin{flushright}$\Box$\end{flushright}}

\baselineskip=20pt

%%%%%%%%%%%%%%%%%%%%%%%%%%%%%%%%%%%%%%%%%%%%%%%%%%%%%%%%%%%%
%                                                          %
%  Title page                                              %
%                                                          %
%%%%%%%%%%%%%%%%%%%%%%%%%%%%%%%%%%%%%%%%%%%%%%%%%%%%%%%%%%%%
\newfont{\elevenmib}{cmmib10 scaled\magstep1}
\newcommand{\preprint}{
   \begin{flushleft}
     %\elevenmib Yukawa\, Institute\, Kyoto\\
   \end{flushleft}\vspace{-1.3cm}
   \begin{flushright}\normalsize
  % \sf  YITP-03-53\\
   %  {\tt hep-th/yymmnnn} \\ November 2005
   \end{flushright}}
\newcommand{\Title}[1]{{\baselineskip=26pt
   \begin{center} \Large \bf #1 \\ \ \\ \end{center}}}
\newcommand{\Author}{\begin{center}
   \large \bf
Xin Zhang${}^{a}$,~Yuan-Yuan Li${}^{a}$,~Junpeng Cao${}^{a,b}$,~Wen-Li Yang${}^{c,d}\footnote{Corresponding author:
wlyang@nwu.edu.cn}$,\\~Kangjie Shi${}^c$ and~Yupeng
Wang${}^{a,b}\footnote{Corresponding author: yupeng@iphy.ac.cn}$
 \end{center}}
\newcommand{\Address}{\begin{center}

     ${}^a$Beijing National Laboratory for Condensed Matter
          Physics, Institute of Physics, Chinese Academy of Sciences, Beijing
           100190, China\\
     ${}^b$Collaborative Innovation Center of Quantum Matter, Beijing,
     China\\
     ${}^c$Institute of Modern Physics, Northwest University,
     Xian 710069, China \\
     ${}^d$Beijing Center for Mathematics and Information Interdisciplinary Sciences, Beijing, 100048,  China
   \end{center}}
\newcommand{\Accepted}[1]{\begin{center}
   {\large \sf #1}\\ \vspace{1mm}{\small \sf Accepted for Publication}
   \end{center}}

\preprint
\thispagestyle{empty}
\bigskip\bigskip\bigskip

\Title{Retrieve the Bethe states of quantum integrable models solved via the off-diagonal Bethe Ansatz} \Author

\Address
\vspace{1cm}

\begin{abstract}
Based on the inhomogeneous $T-Q$ relation constructed via the
off-diagonal Bethe Ansatz, a systematic method for retrieving the
Bethe-type eigenstates of integrable models without obvious reference state is developed
by employing certain orthogonal basis of the Hilbert space. With the
XXZ spin torus model and the open XXX spin-$\frac{1}{2}$ chain as
examples, we show that for a given inhomogeneous $T-Q$ relation and
the associated Bethe Ansatz equations, the constructed Bethe-type
eigenstate has a well-defined homogeneous limit.

\vspace{1truecm} \noindent {\it PACS:} 75.10.Pq, 03.65.Vf, 71.10.Pm

\end{abstract}
\newpage
%%%%%%%%%%%%%%%%%%%%%%%%%%%%%%%%%%%%%%%%%%%%%%%%%%%%%%%%%%%%%%%
%                                                             %
%  1. Introduction                                            %
%                                                             %
%%%%%%%%%%%%%%%%%%%%%%%%%%%%%%%%%%%%%%%%%%%%%%%%%%%%%%%%%%%%%%%
\section{Introduction}
\label{intro} \setcounter{equation}{0}

The algebraic Bethe Ansatz method provides a powerful tool to solve
integrable models with $U(1)$ symmetry \cite{Skl78,Tak79,Skl88,Kor93}.
In that approach, both eigenvalues and eigenstates of
transfer matrix can be constructed simultaneously. However, for
integrable models without $U(1)$ symmetry, Bethe-type
eigenstates can be constructed only for some very special boundary
conditions \cite{Cao03,Yan04-1,Yan04,Doi06,Yan07,Cra12,Gra13,Pim13,Bel133}.
Recently, a new method, namely, the off-diagonal Bethe Ansatz (ODBA)
method \cite{Cao1,Cao2,Cao2-1,Cao3,Cao14} was proposed to approach
exact solutions of generic integrable models either with or
without $U(1)$ symmetry. In such an approach, the spectrum of the
transfer matrix as well as the scalar products between off-shell states and an eigenstate can be observed without using any information of
states. The central point of this method lies in
construction of the inhomogeneous $T-Q$ relation based on
operator product identities. An interesting issue left in this
framework is how to retrieve eigenstates from the obtained spectrum. In
principle, if the eigenvalues of a matrix are known, its
eigenvectors should be determined completely. A remarkable progress in this aspect is the conjecture on the Bethe
states of the open $XXX$ spin chain \cite{Bel13} based on the inhomogeneous $T-Q$ relation.
On the other hand, the eigenstates of
several integrable models in case of inhomogeneity were also derived
via the separation of variables (SoV) method
\cite{Nic12,Nic13,Fad14}. However, how to reach the homogeneous limit of the SoV states and how to prove the conjecture proposed in \cite{Bel13} still remain open.

In this paper, we propose a systematic method to retrieve the
eigenstates from the ODBA solutions. The central point lies in that from the inhomogeneous $T-Q$ relation, one can retrieve the reference state that is normally not known. With this reference state, one can easily reach the homogeneous limit of the Bethe states. We employ two archetype integrable
models without $U(1)$ symmetry, i.e., the $XXZ$ spin torus model and the open $XXX$ spin-$\frac12$ chain with generic boundary fields as examples, to elucidate our method.

The paper is organized as follows. Section 2 serves as an
introduction of our notations and brief review of the ODBA solutions of the inhomogeneous spin
torus. In section 3, after introducing a complete (both left and right) basis of the
Hilbert space, we retrieve the Bethe-type eigenstates of the transfer matrix with the help of the inhomogeneous $T-Q$
relations.
Section 4 is devoted to constructing Bethe states of the open $XXX$ spin-$\frac12$ chain with generic boundary fields. In section 5, we summarize our
results and give the concluding remarks. Some technical proofs are
given in Appendices A and B respectively.

%%%%%%%%%%%%%%%%%%%%%%%%%%%%%%%%%%%%%%%%%%%%%%%%%%%%%%%%%%%%%%%
%                                                             %
%  2. Transfer matrix                                         %
%                                                             %
%                                                             %
%                                                             %
%%%%%%%%%%%%%%%%%%%%%%%%%%%%%%%%%%%%%%%%%%%%%%%%%%%%%%%%%%%%%%%

\section{ODBA Solution of the Spin Torus}
\label{XXZ} \setcounter{equation}{0}
The $XXZ$ spin torus model is described by the Hamiltonian
\begin{eqnarray}
H = -\sum_{n=1}^N(
  \sigma_n^x\sigma_{n+1}^x+\sigma_n^y\sigma_{n+1}^y
  +\cosh\eta\sigma_n^z\sigma_{n+1}^z),\label{xyzh}
\end{eqnarray}
with the anti-periodic boundary condition
\bea
  \sigma^x_{N+1}=\sigma^x_1,\quad   \sigma^y_{N+1}=-\sigma^y_1,\quad
  \sigma^z_{N+1}=-\sigma^z_1.
  \label{Anti-periodic}
\eea The integrability of this model, associated with the following $\bar R$-matrix satisfying the quantum Yang-Baxter equation (QYBE),  has been studied by several
authors \cite{Yun95,Bat95,Gal08,Nie09,Fra11}
\bea
\bar R(u)=\frac{1}{\sinh \eta}\lt(\begin{array}{llll}\sinh(u+\eta)&&&\\&\sinh u&\sinh\eta&\\
&\sinh\eta&\sinh u&\\&&&\sinh(u+\eta)\end{array}\rt),
\label{r-matrix} \eea where the generic complex number $\eta$ is the
crossing parameter.

Let us introduce the ``row-to-row" (or one-row ) monodromy matrix
$T_0(u)$, a $2\times 2$ matrix with operator-valued elements acting
on ${\rm\bf V}^{\otimes N}$, \bea T_0(u)&=&\bar
R_{0N}(u-\theta_N)\bar R_{0\,N-1}(u-\theta_{N-1})\cdots
\bar R_{01}(u-\theta_1)\no\\[6pt]
&=&\left(\begin{array}{cc}\bar A(u)&\bar B(u)\\
\bar C(u)&\bar D(u)\end{array}\right).\label{Mon-V-1} \eea Here
$\{\theta_j|j=1,\cdots,N\}$ are arbitrary free complex parameters
which are usually called inhomogeneous parameters. QYBE implies that
the monodromy matrix given in (\ref{Mon-V-1}) satisfies the
following RTT relation \bea \bar
R_{12}(u-v)\,T_1(u)\,T_2(v)=T_2(v)\,T_1(u)\,\bar
R_{12}(u-v).\label{RLL} \eea Moreover, the corresponding transfer matrix $t(u)$
is given by \bea t(u)=tr_0(\s_0^xT_0(u))=\bar B(u)+\bar
C(u).\label{trans-Periodic} \eea The QYBE and the $Z_2$-symmetry of
the $R$-matrix lead to the fact that the transfer matrices with
different spectral parameters commute with each other:
$[t(u),t(v)]=0$.

The Hamiltonian (\ref{xyzh}) with anti-periodic boundary condition is then expressed in terms of the transfer matrix by
\begin{eqnarray}
H=-2\sinh\eta\left\{\left.\frac{\partial \ln
t(u)}{\partial
u}\right|_{u=0,\{\theta_j=0\}}-\frac{1}{2}N\coth\eta\right\}.\label{ham}
\end{eqnarray}

It was shown  \cite{Cao1} that
an eigenvalue $\Lambda(u)$ of $t(u)$ satisfies the following properties:
\bea
&&\L(u+i\pi)=(-1)^{N-1}\L(u),\label{Eigen-Periodic}\\
&&\Lambda(u) \mbox{, as function of $u$, is a trigonometrical polynomial of degree $N-1$},\label{Eign-anal}\\
&&\L(\theta_j)\,\L(\theta_j-\eta)=-\bar a(\theta_j)\bar
d(\theta_j-\eta),\quad j=1,\cdots, N.\label{Eigen-id} \eea  Here
$\bar a(u)=\prod_{l=1}^N\sinh(u-\theta_l+\eta)$ and $\bar d(u)=\bar
a(u-\eta)=\prod_{l=1}^N\sinh(u-\theta_l)$. The above relations
completely determine the eigenvalue $\L(u)$ as follows \cite{Cao1}:
the eigenvalue $\L(u)$ is given by the following inhomogeneous $T-Q$
relation
\begin{eqnarray}
\Lambda(u)=e^{u}\bar a(u)\frac{Q_1(u-\eta)}{Q_2(u)}
-e^{-u-\eta}\bar d(u)\frac{Q_2(u+\eta)}{Q_1(u)}-c(u)\frac{\bar a(u)\bar d(u)}{Q_1(u)Q_2(u)},
\label{T-Q-1}
\end{eqnarray}
where
\begin{eqnarray}
Q_1(u)=\prod_{j=1}^{M}\frac{\sinh(u-\mu_j)}{\sinh\eta},\quad
Q_2(u)=\prod_{j=1}^{M}\frac{\sinh(u-\nu_j)}{\sinh\eta},\label{Q-1}
\end{eqnarray}
and $c(u)$ is an adjust function. For an even $N$ and $M=\frac{N}{2}$, $c(u)$ is given by
\begin{eqnarray}
c(u)=e^{u+\sum_{l=1}^N\theta_l-M\eta-2\sum_{j=1}^M\mu_j}-e^{-u-\eta-\sum_{l=1}^N\theta_l+M\eta+2\sum_{j=1}^M\nu_j},\label{C-1}
\end{eqnarray}
where the $N$ parameters $\{\mu_j\}$ and $\{\nu_j\}$ satisfy the following Bethe Ansatz equations (BAEs)
\begin{eqnarray}
&&\bar d(\nu_j)=\frac{e^{\nu_j}}{c(\nu_j)}Q_1(\nu_j-\eta)Q_1(\nu_j),\quad \quad j=1,\cdots, M,\label{BAE-1}\\
&&\bar
a(\mu_j)=-\frac{e^{-\mu_j-\eta}}{c(\mu_j)}Q_2(\mu_j+\eta)Q_2(\mu_j),\quad
j=1,\cdots, M.\label{BAE-2}
\end{eqnarray}
For an odd $N$ and $M=\frac{N+1}{2}$, $c(u)$ is given by
\begin{eqnarray}
c(u)=\frac{1}{2\sinh\eta}\lt\{e^{2u+\sum_{l=1}^N\theta_l-M\eta-2\sum_{j=1}^M\mu_j}+e^{-2u-2\eta-\sum_{l=1}^N\theta_l+M\eta+2\sum_{j=1}^M\nu_j}\rt\},\label{C-2}
\end{eqnarray}
and the $N+1$ parameters $\{\mu_j\}$ and $\{\nu_j\}$ satisfy the BAEs (\ref{BAE-1})-(\ref{BAE-2}) with the adjust function $c(u)$ given by (\ref{C-2}).

It was known in \cite{Cao2} that there actually exist many different types of $T-Q$ relations for the solutions to (\ref{Eigen-Periodic})-(\ref{Eigen-id})
and each of them gives the complete set of eigenvalues of the transfer matrix as proven in \cite{Cao14-2}. Here we present another simple $T-Q$ relation for $\L(u)$, which corresponds
to the $M=0$ type  in \cite{Cao2}, namely,
\begin{eqnarray}
\Lambda(u)=\bar a(u)e^{u}\frac{Q(u-\eta)}{Q(u)}-e^{-u-\eta}\bar d(u)\frac{Q(u+\eta)}{Q(u)}-c(u)\frac{\bar a(u)\bar d(u)}{Q(u)},\label{T-Q-2}
\end{eqnarray}
with
\begin{eqnarray}
Q(u)=\prod_{j=1}^N\frac{\sinh(u-\lambda_j)}{\sinh\eta},\label{Q-2}
\end{eqnarray}
and
\begin{eqnarray}
c(u)=e^{u-N\eta+\sum_{j=1}^N(\theta_j-\lambda_j)}-e^{-u-\eta-\sum_{j=1}^N(\theta_j-\lambda_j)}. \label{C-3}
\end{eqnarray}
The $N$ parameters $\{\l_j\}$ in (\ref{Q-2}) satisfy the associated BAEs
\begin{eqnarray}
\hspace{-0.8truecm}e^{\lambda_j}\bar a(\lambda_j)Q(\lambda_j-\eta)-
\bar
d(\lambda_j)e^{-\lambda_j-\eta}Q(\lambda_j+\eta)-c(\lambda_j)\bar
a(\lambda_j)\bar d(\lambda_j)=0, \quad j=1,\cdots,N.\label{BAE-3}
\end{eqnarray}

%%%%%%%%%%%%%%%%%%%%%%%%%%%%%%%%%%%%%%%%%%%%%%%%%%%%%%%%%%%%%%%
%                                                             %
%  4. Retrieving Eigenstates                                  %
%                                                             %
%                                                             %
%                                                             %
%%%%%%%%%%%%%%%%%%%%%%%%%%%%%%%%%%%%%%%%%%%%%%%%%%%%%%%%%%%%%%%

\section{ Retrieving the Eigenstates  }
\label{BAs} \setcounter{equation}{0}

With generic inhomogeneous parameters $\theta_j\neq\theta_l$ and
$\theta_j\neq \theta_l-\eta$, a set of orthogonal
states parameterized by the $N$ inhomogeneous constants
$\{\theta_j|j=1,\cdots,N\}$ that form a basis of the Hilbert space exists. In the framework of ODBA, such a
basis is quite useful to prove the retrieved Bethe states to be
eigenstates of the transfer matrix.

\subsection{Orthogonal basis}

The RTT relation (\ref{RLL}) of the monodromy matrix $T(u)$ given by
(\ref{Mon-V-1}) gives rise to some quadratic commutation relations
among its matrix elements. Here we present some relevant ones for our
purpose
\begin{eqnarray}
&&[\bar B(u),\bar B(v)]=[\bar C(u),\bar C(v)]=0, \label{RLL-1} \\[4pt]
&& \bar A(u)\bar B(v)=\frac{\sinh(u-v-\eta)}{\sinh(u-v)}\bar B(v)
\bar A(u)+\frac{\sinh\eta}{\sinh(u-v)}\bar B(u)\bar A(v), \label{RLL-2} \\[4pt]
&& \bar D(u)\bar B(v)=\frac{\sinh(u-v+\eta)}{\sinh(u-v)}\bar B(v)\bar D(u)
-\frac{\sinh\eta}{\sinh(u-v)}\bar B(u)\bar D(v), \label{RLL-3} \\[4pt]
&& \bar C(u)\bar A(v)=\frac{\sinh(u-v+\eta)}{\sinh(u-v)}\bar A(v)
\bar C(u)-\frac{\sinh\eta}{\sinh(u-v)}\bar A(u)\bar C(v), \label{RLL-4} \\[4pt]
&& \bar C(u)\bar D(v)=\frac{\sinh(u-v-\eta)}{\sinh(u-v)}\bar D(v)
\bar C(u)+\frac{\sinh\eta}{\sinh(u-v)}\bar D(u)\bar C(v), \label{RLL-5} \\[4pt]
&& [\bar C(u),\bar B(v)]=\frac{\sinh\eta}{\sinh(u-v)}[\bar D(u)\bar A(v)-\bar D(v)\bar A(u)]. \label{RLL-6}
\end{eqnarray}
Let us introduce the all spin up state $|0\rangle$  and its dual state $\langle 0|$
\bea
|0\rangle=\otimes_{j=1}^N|\uparrow\rangle_j,\quad \langle 0|=\langle \uparrow|_{j}\otimes_{j=1}^N,\label{Dual-up}
\eea
which are nothing but the reference state and its dual in the framework of
the algebraic Bethe Ansatz method \cite{Kor93}. The elements of the monodromy matrix
act on them as follows:
\bea
&&\bar A(u)|0\rangle=\bar a(u)|0\rangle,\quad \bar D(u)|0\rangle=\bar d(u)|0\rangle,\quad \bar C(u)|0\rangle=0,\label{Ref-1}\\
&&\langle 0| \bar A(u)= \bar a(u)\langle 0|,\quad \langle 0|\bar D(u)=\bar d(u)\langle 0|,\quad \langle 0|\bar B(u)=0.\label{Ref-2}
\eea

Let us introduce some left and right states parameterized by the $N$ inhomogeneous parameters $\{\theta_j\}$
as follows \footnote{These states were used in \cite{Nic12} to construct
Sklyanin's quantum separation of variables (SoV) \cite{Skl92}
representations of the Yang-Baxter algebra associated with the
trigonometric six-vertex $R$-matrix.}:
\begin{eqnarray}
\langle\theta_{p_1},\cdots,\theta_{p_n}|&=&\langle 0|\prod_{j=1}^n\bar C(\theta_{p_j}),\quad 1\leq p_1<p_2<\cdots<p_n\leq N,\label{left-state}\\[4pt]
|\theta_{q_1},\cdots,\theta_{q_n}\rangle&=&\prod_{j=1}^n\bar
B(\theta_{q_j})|0\rangle,\quad 1\leq q_1<q_2<\cdots<q_n\leq
N.\label{right-state}
\end{eqnarray}
Due to the fact that $\bar d(\theta_j)=0$, with the help  of (\ref{RLL-3}) and (\ref{RLL-5}) one may derive that these
states  are in fact the eigenstates of $\bar D(u)$
\begin{eqnarray}
\bar D(u)|\theta_{p_1},\cdots,\theta_{p_n}\rangle=\bar
d(u)\prod_{j=1}^n\frac{\sinh(u-\theta_{p_j}+\eta)}{\sinh(u-\theta_{p_j})}|\theta_{p_1},
\cdots,\theta_{p_n}\rangle,\label{Eigen-D-1}\\[4pt]
\langle\theta_{p_1},\cdots,\theta_{p_n}|\bar D(u)=\bar
d(u)\prod_{j=1}^n\frac{\sinh(u-\theta_{p_j}+\eta)}{\sinh(u-\theta_{p_j})}
\langle\theta_{p_1},\cdots,\theta_{p_n}|.\label{Eigen-D-2}
\end{eqnarray}
Note that the total number of the right (or left) states given in (\ref{right-state}) (or (\ref{left-state})) is
\begin{eqnarray}
\sum_{n=0}^N\frac{N!}{(N-n)!n!}=2^N.\no
\end{eqnarray}

Using the commutation relations (\ref{RLL-1})-(\ref{RLL-6}), one may derive the following orthogonal relations between the left states and the right states
\begin{eqnarray}
\langle\theta_{p_1},\cdots,\theta_{p_n}|\theta_{q_1},\cdots,\theta_{q_m}\rangle
=f_n(\theta_{p_1},\cdots,\theta_{p_n})\,\delta_{m,n}\,\prod_{j=1}^n\delta_{p_j,q_j},\label{Orth-relation-1}
\end{eqnarray}
where
\begin{eqnarray}
\hspace{-0.8truecm}f_n(\theta_{p_1},\cdots,\theta_{p_n})
=\prod_{l=1}^n\lt\{\bar a(\theta_{p_l})\bar d_{p_l}(\theta_{p_l})
\prod_{k\neq
l}^n\frac{\sinh(\theta_{p_l}-\theta_{p_k}+\eta)}{\sinh(\theta_{p_l}-\theta_{p_k})}\rt\}.
\label{Orth-relation-2}
\end{eqnarray} We remark that $f_0=\langle0|0\rangle=1$. The function ${\bar d}_l(u)$ is defined as
\bea \bar d_l(u)=\prod_{j\neq
l}^N\frac{\sinh(u-\theta_j)}{\sinh\eta},\quad l=1,\cdots,N.
\label{d-function} \eea Thus for generic values $\{\theta_j\}$,
these right (or left) states form an orthogonal right (or left)
basis of the Hilbert space, and any right (or left) state can be
decomposed as a unique linear combination of these basis.

\subsection{Retrieving the Bethe states}

Due to the fact that  the left states
$\{\langle\theta_{p_1},\cdots,\theta_{p_n}|\,|n=0,\cdots,N,\,\quad
1\leq p_1<p_2<\cdots<p_n\leq N\}$ given by (\ref{left-state}) form a
basis of the dual Hilbert space, an eigenstate $|\Psi\rangle$ of the transfer matrix (\ref{trans-Periodic}) is completely determined (up to an
overall scalar factor) by the following set of scalar products \bea
 F_n(\theta_{p_1},\cdots,\theta_{p_n})=\langle\theta_{p_1},\cdots,\theta_{p_n}|\Psi\rangle,\quad n=0,\cdots,N,\label{F-functions-XXZ}
\eea with the first one $F_0=1$. It was shown  in \cite{Cao1} that these products 
$F_n(\theta_{p_1},\cdots,\theta_{p_n})$  are given by \bea
F_n(\theta_{p_1},\cdots,\theta_{p_n})=\prod_{l=1}^n\L(\theta_{p_l}),\quad
n=1,\cdots,N.\label{Fn-vaules} \eea

Let us consider the following Bethe state
\begin{eqnarray}
|\lambda_1,\cdots,\lambda_N\rangle=\prod_{j=1}^N\frac{\bar
D(\lambda_j)}{\bar d(\lambda_j)}\,|\Omega;\{\theta_j\}
\rangle,\label{Bethe-state-1}
\end{eqnarray}
where the parameters $\{\lambda_j|j=1,\cdots,N\}$ satisfy the BAEs
(\ref{BAE-3}) and  $|\Omega;\{\theta_j\}\rangle$ is a generalized
reference state to be determined so that the scalar products between the Bethe state (\ref{Bethe-state-1})
and the basis (\ref{left-state}) satisfy the conditions
(\ref{Fn-vaules}), namely,
\bea
\langle\theta_{p_1},\cdots,\theta_{p_n}| \lambda_1,\cdots,\lambda_N\rangle=
\prod_{l=1}^n\L(\theta_{p_l}),\quad
n=1,\cdots,N.\label{Fn-vaules-1}
\eea
For an eigenvalue $\L(u)$ given by  the
$T-Q$ relation (\ref{T-Q-2}), its value at the inhomogeneous point
$\theta_j$ takes a simple form: \bea \L(\theta_j)=\bar
a(\theta_j)\,e^{\theta_j}\frac{Q(\theta_j-\eta)}{Q(\theta_j)},\quad
j=1,\cdots,N.\label{Simple-form} \eea  With the help of  the
exchange relation (\ref{RLL-5}) and the above relations, the
conditions (\ref{Fn-vaules-1}) are then equivalent to the following
requirements on the reference state: \bea \langle
\theta_{p_1},\cdots,\theta_{p_n}|
\Omega;\{\theta_j\}\rangle=\prod_{l=1}^na(\theta_{p_l})e^{\theta_{p_l}},\quad
n=0,\cdots,N,\,\,1\leq p_1<p_2<\cdots<p_n\leq
N.\label{refernece-conditions} \eea
It is remarked that the above conditions do not depend on the
parameters $\{\lambda_j|j=1,\cdots,N\}$ which should satisfy the BAEs (\ref{BAE-3}).
Hence the relations (\ref{refernece-conditions})
uniquely determine the reference state $|
\Omega;\{\theta_j\}\rangle$ up to a scalar factor, which actually does not depend upon the parameters
$\{\lambda_j|j=1,\cdots,N\}$.

Let us propose the following Ansatz for the reference state $|\Omega;\{\theta_j\}\rangle$:
\bea
|\Omega;\{\theta_j\}\rangle=\sum_{l=0}^{\infty}\frac{\lt(\tilde{B}^{-}\rt)^l}{[l]_q!}|0\rangle
=\sum_{l=0}^{N}\frac{\lt(\tilde{B}^{-}\rt)^l}{[l]_q!}|0\rangle,\label{reference-state}
\eea where the $q$-integers $\{[l]_q|l=0,\cdots\}$ and the operator $\tilde{B}^{-}$ are given by
\bea
&&[l]_q=\frac{1-q^{2l}}{1-q^{2}},\quad [0]_q=1,\\
&&[l]_q!=[l]_q\,[l-1]_q\cdots [1]_q,\quad q=e^{\eta},\\[4pt]
&&\tilde{B}^{-}=\lim_{u\to +\infty}\lt\{\lt(2\sinh\eta\,e^{-u}\rt)^{N-1}e^{\sum_{l=1}^N\theta_l}\,\bar B(u)\rt\}. \label{4-limit}
\eea The definitions (\ref{r-matrix}) and (\ref{Mon-V-1}) allow us to obtain the explicit expression of the operator $\tilde{B}^{-}$ as
\bea
\tilde{B}^{-}=\sum_{l=1}^Ne^{\theta_l+\frac{(N-1)\eta}{2}}\,e^{\frac{\eta}{2}\sum_{k=l+1}^N\sigma^z_k}\,\sigma^-_l \,
e^{-\frac{\eta}{2}\sum_{k=1}^{l-1}\sigma^z_k}.\label{B-operator-1}
\eea
Direct calculation shows that the state $|\Omega;\{\theta_j\}\rangle$ given by (\ref{reference-state}) indeed satisfies the relations
(\ref{refernece-conditions}). The proof is given in Appendix A.  Then we conclude that the Bethe state (\ref{Bethe-state-1})
with the corresponding reference state (\ref{reference-state})
is an eigenstate of the transfer matrix (\ref{trans-Periodic}), provided that the parameters $\{\lambda_j|j=1,\cdots,N\}$ satisfy the associated BAEs
(\ref{BAE-3}). The corresponding eigenvalue is given by the $T-Q$ relation (\ref{T-Q-2}).

From the definitions (\ref{r-matrix}) and (\ref{Mon-V-1}), one can see that the operators $\bar D(u)$ and $\tilde{B}^{-}$ have
well-defined homogeneous limit when $\{ \theta_j\rightarrow 0\}$.
In the homogeneous limit, the reference state (\ref{reference-state}) becomes
\begin{eqnarray}
|\Omega\rangle=\lim_{\{\theta_j\to 0\}}|\Omega; \{\theta_j\}\rangle
=\sum_{l=0}^{\infty}\frac{\lt(B^{-}\rt)^l}{[l]_q!}|0\rangle
=\sum_{l=0}^{N}\frac{\lt(B^{-}\rt)^l}{[l]_q!}|0\rangle
,\label{Reference-state-2}
\end{eqnarray} where the operator $B^{-}$ (c.f., (\ref{B-operator-1})) reads
\bea
B^-=\lim_{\{\theta_j\to 0\}}\,\tilde{B}^-=
\sum_{l=1}^Ne^{\frac{(N-1)\eta}{2}}\,e^{\frac{\eta}{2}\sum_{k=l+1}^N\sigma^z_k}\,\sigma^-_l \,
e^{-\frac{\eta}{2}\sum_{k=1}^{l-1}\sigma^z_k}.
\eea
This implies that the homogeneous limit of the Bethe state (\ref{Bethe-state-1})
gives rise to  the eigenstate\footnote{It is remarked that in contrast with
the Hamiltonian given by  (\ref{xyzh}) and (\ref{Anti-periodic}),
the reference state (\ref{Reference-state-2})  is not invariant
under the translation. This is due to that the reference state is no longer
an eigenstate of the Hamiltonian. } of
the corresponding homogeneous transfer matrix.
The corresponding eigenvalue and BAEs are given by  the homogeneous limits of (\ref{T-Q-2}) and (\ref{BAE-3}) respectively.  It should be noted that in contrast to that used in the algebraic Bethe Ansatz scheme, the reference state  (\ref{reference-state}) (or (\ref{Reference-state-2})) is no longer a pure product state
but a highly entangled state (actually a $q$-spin coherent state).

Associated with the $T-Q$ relation (\ref{T-Q-1}), we can construct another type of Bethe states
\begin{eqnarray}
|\mu_1,\cdots,\mu_M;\nu_1,\cdots,\nu_M\rangle=\prod_{j=1}^M\frac{\bar
D(\mu_j)}{\bar d(\mu_j)}\frac{\bar D(\nu_j)}{\bar d(\nu_j)}
|\bar\Omega; \{\theta_j\}\rangle,\label{Bethe-state-2}
\end{eqnarray}
where the associated reference state is
\begin{eqnarray}
\hspace{-0.6truecm} |\bar\Omega;
\{\theta_j\}\rangle=\sum_{n=0}^N\sum_{p}f_n^{-1}(\theta_{p_1},\cdots,\theta_{p_n})\prod_{l=1}^n
e^{\theta_{p_l}}\bar
a(\theta_{p_l})\frac{Q_1(\theta_{p_l})}{Q_2(\theta_{p_l}-\eta)}|\theta_{p_1},\cdots,\theta_{p_n}\rangle,\label{Reference-state-3}
\end{eqnarray} with the $Q-$functions $Q_1(u)$ and $Q_2(u)$ given by (\ref{Q-1}).
It can be easily checked that
\begin{eqnarray}
\hspace{-0.6truecm}\langle\theta_{p_1},\cdots,\theta_{p_n}|\mu_1,\cdots,\mu_M;\nu_1,\cdots,\nu_M\rangle&=&
\prod_{l=1}^n\,e^{\theta_{p_l}}\bar a(\theta_{p_l})\frac{Q_1(\theta_{p_l}-\eta)}{Q_2(\theta_{p_l})}\nonumber\\[4pt]
\hspace{-0.6truecm}&=&\prod_{l=1}^n\Lambda(\theta_{p_l})=F_n(\theta_{p_1},\cdots,\theta_{p_n}).\no
\end{eqnarray}
Therefore, the Bethe state (\ref{Bethe-state-2}) is also an
eigenstate of the transfer matrix provided that the parameters
$\{\mu_j\}$ and $\{\nu_j\}$ satisfy the associated BAEs
(\ref{BAE-1})-(\ref{BAE-2}). In the homogeneous limit, the reference state (\ref{Reference-state-3}) reads
\begin{eqnarray}
|\bar\Omega\rangle=\lim_{\{\theta_j\to 0\}}|\bar \Omega; \{\theta_j\}\rangle=\sum_{l=0}^{\infty}\frac{\lt({Q_1(0)B^{-}}\rt)^l}{[l]_q!\lt({Q_2(-\eta)}\rt)^l}|0\rangle
=\sum_{l=0}^{N}\frac{\lt({Q_1(0)B^{-}}\rt)^l}{[l]_q!\lt({Q_2(-\eta)}\rt)^l}|0\rangle
.\label{Reference-state-4}
\end{eqnarray}

Some remarks are in order. Different choice of the inhomogeneous
$T-Q$ relations gives different parameterization of the eigenvalues
and  leads to a different expression of the Bethe state and the
associated reference state\footnote{There exists a homogeneous $T-Q$ relation
for  $\Lambda(u)$ \cite{Bat95}. However, how to retrieve the corresponding Bethe
state for such a parametrization is still an open and interesting problem.}. The procedure for constructing the Bethe
states (\ref{Bethe-state-1}) and (\ref{Bethe-state-2}) is different
from that of the algebraic Bethe Ansatz. In the latter scheme, one
uses known reference state and creation operator to derive
eigenvalues and eigenstates of the transfer matrix, while in the
ODBA scheme one uses known eigenvalues (in terms of inhomogeneous
$T-Q$ relation)  and creation operator  to retrieve the reference
state.  The key point is that the eigenstates of the creation
operator (i.e., $\bar D(u)$)  form a basis of the Hilbert space.
Such a reversed process makes it convenient to approach the
eigenstate problem of quantum integrable models without obvious
reference state.

%%%%%%%%%%%%%%%%%%%%%%%%%%%%%%%%%%%%%%%%%%%%%%%%%%%%%%%%%%%%%%%
%                                                             %
%  6. Results for the XXX spin-$\frac{1}{2}$ open chain       %
%                                                             %
%                                                             %
%                                                             %
%%%%%%%%%%%%%%%%%%%%%%%%%%%%%%%%%%%%%%%%%%%%%%%%%%%%%%%%%%%%%%%

\section{ Results for the open XXX spin-$\frac{1}{2}$ chain}
\label{XXXX open chain} \setcounter{equation}{0}

In this section, we show how to retrieve the Bethe states conjectured in \cite{Bel13} from the ODBA solution of the open XXX spin-$\frac{1}{2}$ chain described by the Hamiltonian
\begin{eqnarray}
H=\sum_{j=1}^{N-1}{\vec \sigma}_{j}\cdot {\vec \sigma}_{j+1}+\frac{\eta}{p}
\sigma_{1}^{z} +\frac{\eta}{q}( \xi\sigma_{N}^{x}+ \sigma_{N}^{z}),
\label{Open-XXX}
\end{eqnarray}
where $p$, $q$ and $\xi$ are  arbitrary boundary parameters. The corresponding transfer matrix of the
inhomogeneous open chain is given by \cite{Skl88}
\bea
t^{(o)}(u)=tr_0\lt(K^+_0(u)T_0(u)K^-_0(u)\hat{T}_0(u)\rt),\label{Trans-XXX-open}
\eea
with the monodromy matrices $T(u)$ and $\hat{T}(u)$ defined as
\bea
T_0(u)&=&R_{0N}(u-\theta_N)\cdots R_{01}(u-\theta_1)=\left(\begin{array}{cc} A(u)& B(u)\\
 C(u)& D(u)\end{array}\right),\label{Mon-XXX-1}\\[6pt]
\hat{T}_0(u)&=&R_{01}(u\hspace{-0.08truecm}+\hspace{-0.08truecm}\theta_1)\cdots
R_{0N}(u\hspace{-0.08truecm}+\hspace{-0.08truecm}\theta_N)
 =(-1)^N\left(\begin{array}{cc} D(-u-\eta)& -B(-u-\eta)\\
 -C(-u-\eta)& A(-u-\eta)\end{array}\right),\label{Mon-XXX-2}
\eea where the associated $R$-matrix $R(u)$ (c.f. (\ref{r-matrix})) reads
\bea
R(u)=\lt(\begin{array}{llll}u+\eta&&&\\&u&\eta&\\
&\eta& u&\\&&&u+\eta\end{array}\rt),
\label{r-matrix-1} \eea
and the  $K$-matrices are given by \cite{Veg93, Gho94}
\begin{eqnarray}
K^-(u)&=&\left(\begin{array}{cc}
p+u & 0\\
0 & p-u
\end{array}\right)
\stackrel{{\rm def}}{=}\lt(\begin{array}{cc}
 K^-_{11}(u)& K^-_{12}(u)\\
 K^-_{21}(u)& K^-_{22}(u)\\
\end{array}\rt)
,\label{K-matrices-1}\\[6pt]
K^+(u)&=&\left(\begin{array}{cc}
q+u+\eta & \xi(u+\eta)\\
\xi(u+\eta) & q-u-\eta
\end{array}\right)
\stackrel{{\rm def}}{=}\lt(\begin{array}{cc}
 K^+_{11}(u)& K^+_{12}(u)\\
 K^+_{21}(u)& K^+_{22}(u)\\
\end{array}\rt).\label{K-matrices-2}
\end{eqnarray}
The transfer matrix has the commutative property $[t(u), t(v)] = 0$,
and is therefore the generating functional of a family of commuting
operators, among which is the Hamiltonian (\ref{Open-XXX}), i.e.,
\bea H=\eta\frac{\partial \ln t^{(o)}(u)}{\partial
u}|_{u=0,\theta_j=0}-N.\nonumber \eea

%%%%%%%%%%%%%%%%%%%%%%%%%%%%%%%%%%%%%%%%%%%%%%%%%%%%%%%%%%%%%%%%%%%%%%%%%%%%%%%%%%%%

\subsection{ ODBA solution and the associated basis}

Let us introduce the following functions
\bea
a(u)&=&\prod_{l=1}^N(u-\theta_l+\eta),\quad d(u)=a(u-\eta)=\prod_{l=1}^N(u-\theta_l).\label{a-b-functions-rational}
\eea
Each eigenvalue of the transfer matrix (\ref{Trans-XXX-open}), denoted by $ \Lambda^{(o)}(u)$, can be  given in terms of the following
inhomogeneous $T-Q$ relation \cite{Cao2,Nep13,Cao14-2} \footnote{The $T-Q$ relation (\ref{T-Q-XXX}) corresponds to 
the  special case (i.e., $M=0$) of the general ones proposed in \cite{Cao2}.  A generalization to other cases is straightforward. }
\bea
 \Lambda^{(o)}(u)&=&(-1)^{N}\frac{2u+2\eta}{2u+\eta}(u+ p)(\sqrt{1+\xi^2}\,u+ q)a(u)d(-u-\eta)\frac{Q(u-\eta)}{Q(u)} \no\\
&&+(-1)^{N}\frac{2u}{2u+\eta}(u- p+\eta)(\sqrt{1+\xi^2}\,(u+\eta)- q)a(-u-\eta)d(u)\frac{Q(u+\eta)}{Q(u)}\no\\
&&+2(1-\sqrt{1+\xi^2})u(u+\eta)\frac{a(u)a(-u-\eta)d(u)d(-u-\eta)}{Q(u)},\label{T-Q-XXX}
\eea where the $Q$-function is given by
\bea
Q(u)=\prod_{j=1}^N(u-\l_j)(u+\l_j+\eta).\label{Q-function-XXX}
\eea The parameters $\{\l_j\}$ satisfy the following BAEs
\bea
&&1+\frac{\l_j\,(\l_j- p+\eta)\,(\sqrt{1+\xi^2}\,(\l_j+\eta)- q)\,a(-\l_j-\eta)\,d(\l_j)\,Q(\l_j+\eta)}
{(\l_j+\eta)\,(\l_j+ p)\,(\sqrt{1+\xi^2}\,\l_j+ q)\,a(\l_j)\,d(-\l_j-\eta)\,Q(\l_j-\eta)}\no\\
&&\quad\quad
=(-1)^N\frac{(\sqrt{1+\xi^2}-1)\,\l_j\,(2\l_j+\eta)\,a(-\l_j-\eta)\,d(\l_j)}
{(\l_j+ p)\,(\sqrt{1+\xi^2}\,\l_j+ q)\,Q(\l_j-\eta)}, \quad
j=1,\cdots,N.\label{BAE-XXX} \eea

It is easy to check that the $K^+$-matrix can be diagonalized as
\bea
\bar{K}^+(u)&=&UK^+(u)U^{-1}=\left(
                             \begin{array}{cc}
                               q+\sqrt{1+\xi^2}(u+\eta) & 0 \\
                               0 & q-\sqrt{1+\xi^2}(u+\eta) \\
                             \end{array}
                           \right)\no\\[6pt]
&\stackrel{{\rm def}}{=}&\lt(\begin{array}{cc}
\bar K^+_{11}(u)&0\\
0&\bar K^+_{22}(u)\\
\end{array}\rt),\label{bar-K-1}
\end{eqnarray} where the matrix $U$ is given by
\bea
U=\left(
\begin{array}{cc}
\xi & \sqrt{1+\xi^2}-1 \\
\xi & -\sqrt{1+\xi^2}-1 \\
\end{array}
\right). \label{U-matrix}
\eea Accordingly, the gauged  $K$-matrix $\bar K^-(u)$ reads
\bea
\bar{K}^-(u)&=&UK^-(u)U^{-1}=\left(
               \begin{array}{cc}
                 p+\frac{1}{\sqrt{1+\xi^2}}u & \frac{\sqrt{1+\xi^2}-1}{\sqrt{1+\xi^2}}u \\
                 \frac{\sqrt{1+\xi^2}+1}{\sqrt{1+\xi^2}}u & p-\frac{1}{\sqrt{1+\xi^2}}u \\
               \end{array}
             \right)\no\\[6pt]
&\stackrel{{\rm def}}{=}&\lt(\begin{array}{cc}
\bar K^-_{11}(u)&\bar K^-_{12}(u)\\
\bar K^-_{21}(u)&\bar K^-_{22}(u) \\
\end{array}\rt).\label{bar-K-2}
\eea Moreover, let us introduce two local states
\begin{eqnarray}
|1\rangle_n&=&\frac{\sqrt{1+\xi^2}+1}{2\xi\sqrt{1+\xi^2}}|\uparrow\rangle_n+\frac{1}{2\sqrt{1+\xi^2}}|\downarrow\rangle_n,
\quad n=1,\cdots,N, \label{State-1}\\[6pt]
|2\rangle_n&=&\frac{\sqrt{1+\xi^2}-1}{2\xi\sqrt{1+\xi^2}}|\uparrow\rangle_n-\frac{1}{2\sqrt{1+\xi^2}}|\downarrow\rangle_n,
\quad n=1,\cdots,N,\label{State-2}
\end{eqnarray}
and their dual states
\begin{eqnarray}
\langle1|_n=\xi\langle\uparrow|_n\hspace{-0.08truecm}+\hspace{-0.08truecm}
(\sqrt{1\hspace{-0.08truecm}+\hspace{-0.08truecm}\xi^2}-\hspace{-0.08truecm}1)\langle\downarrow|_n,\,\,
\langle2|_n=\xi\langle\uparrow|_n\hspace{-0.08truecm}-\hspace{-0.08truecm}(\sqrt{1\hspace{-0.08truecm}+\hspace{-0.08truecm}\xi^2}+\hspace{-0.08truecm}1)\langle\downarrow|_n,\,
n=1,\cdots,N.\label{Dual-state}
\end{eqnarray} These states satisfy the following orthogonal relations
\begin{eqnarray}
\langle a|_jb\rangle_k=\delta_{a,b}\,\delta_{j,k},\quad
a,b=1,2,\quad j,k=1,\cdots,N.\no
\end{eqnarray}
Based on the above local states, let us introduce two product  states
\bea
|\Omega\rangle_{\xi}=\otimes_{j=1}^N|1\rangle_j,\quad _{\xi}\langle\bar{\Omega}|=\otimes_{j=1}^N\langle2|_j .\label{Guage-Ref}
\eea The double-row monodromy matrix of the present model reads
\bea
\mathbb{T}(u)=T(u)\,K^-(u)\,\hat{T}(u)=\lt(\begin{array}{cc}{\cal{A}}(u)& {\cal{B}}(u)\\
{\cal{C}}(u)&{\cal{D}}(u)
\end{array}\rt),\label{2-row-Mon-1}
\eea and its gauged one is
\bea
\bar \mathbb{T}(u)&=&U\,T(u)\,K^-(u)\,\hat{T}(u)U^{-1}=UT(u)U^{-1}\,UK^-(u)U^{-1}\,U\hat{T}(u)U^{-1}\no\\[6pt]
&=&\bar T(u)\,\bar K^-(u) \hat{\bar{T}}(u)=
\lt(\begin{array}{cc}\bar {\cal{A}}(u)& \bar {\cal{B}}(u)\\
\bar {\cal{C}}(u)&\bar {\cal{D}}(u)
\end{array}\rt).\label{2-row-Mon-2}
\eea The double-row monodromy matrix and its gauged one both satisfy
the reflection algebra \cite{Skl88} and the exchange relations among
${\cal{A}}(u)$, ${\cal{B}}(u)$, ${\cal{C}}(u)$ and ${\cal{D}}(u)$
are the same as those among $\bar {\cal{A}}(u)$, $\bar
{\cal{B}}(u)$, $\bar {\cal{C}}(u)$ and $\bar {\cal{D}}(u)$, which
are listed in Appendix B. The transfer matrix $t^{(o)}(u)$ given by
(\ref{Trans-XXX-open}) can be expressed as \bea
t^{(o)}(u)&=&K^+_{11}(u)\,{\cal{A}}(u)+K^+_{12}(u)\,{\cal{C}}(u)+K^+_{21}(u)\,{\cal{B}}(u)+K^+_{22}(u)\,{\cal{D}}(u)\no\\
&=&\bar K^+_{11}(u)\,\bar {\cal{A}}(u)+\bar K^+_{22}(u)\,\bar {\cal{D}}(u). \label{Trans-XXX-1}
\eea

Noting that $\bar {\cal{C}}(u)$ forms a commuting family, i.e., $[\bar {\cal{C}}(u), \bar {\cal{C}}(v)]=0$, similarly as (\ref{left-state})-(\ref{right-state}) we can use its common (dual) eigenstates to construct the basis of right (left)
Hilbert space. For this purpose, let us introduce the following right and left states
parameterized by the $N$ inhomogeneous parameters $\{\theta_j\}$\footnote{Similar basis was used in \cite{Nic13}, where the open $XXZ$ spin chain was studied.}:
\begin{eqnarray}
|\theta_{p_1},\cdots,\theta_{p_n}\rangle\rangle&=&\bar
{\cal{A}}(\theta_{p_1})\cdots\bar {\cal{A}}(\theta_{p_n})
|\Omega\rangle_{\xi},\quad 1\leq p_1<p_2<\cdots<p_n\leq N,\label{Open-right-state}\\
\langle\langle-\theta_{q_1},\cdots,-\theta_{q_n}|&=&_{\xi}\langle\bar{\Omega}|
\bar {\cal{D}}(-\theta_{q_1})\cdots\bar
{\cal{D}}(-\theta_{q_n}),\quad 1\leq q_1<q_2<\cdots<q_n\leq N.
\label{Open-left-state}
\end{eqnarray}
which are eigenstates of $\bar {\cal{C}}(u)$
\begin{eqnarray}
\bar {\cal{C}}(u)|\theta_{p_1},\cdots,\theta_{p_n}\rangle\rangle&=&
h(u,\{\theta_{p_1},\cdots,\theta_{p_n}\})|\theta_{p_1},\cdots,\theta_{p_n}\rangle\rangle,\label{C-eigenstate-1}\\
\langle\langle-\theta_{p_1},\cdots,-\theta_{p_n}|\bar
{\cal{C}}(u)&=&
h'(u,\{-\theta_{p_1},\cdots,-\theta_{p_n}\})\langle\langle-\theta_{p_1},\cdots,-\theta_{p_n}|,\label{C-eigenstate-2}
\end{eqnarray}
with the corresponding eigenvalues being
\begin{eqnarray}
h(u,\{\theta_{p_1},\cdots,\theta_{p_n}\})=(-1)^N\bar
K^-_{21}(u)d(u)d(-u-\eta)\prod_{j=1}^n\frac{(u+\theta_{p_j})(u-\theta_{p_j}+\eta)}
{(u-\theta_{p_j})(u+\theta_{p_j}+\eta)},\label{C-eigenvalue-1}\\[2pt]
h'(u,\{-\theta_{p_1},\cdots,-\theta_{p_n}\})=(-1)^N\bar
K^-_{21}(u)a(u)a(-u-\eta)\prod_{j=1}^n\frac{(u-\theta_{p_j})(u+\theta_{p_j}+\eta)}
{(u+\theta_{p_j})(u-\theta_{p_j}+\eta)}.\label{C-eigenvalue-2}
\end{eqnarray}

For generic inhomogeneous parameters $\{\theta_j\}$, the above relations imply that the left states and right states
satisfy the following relations
\begin{eqnarray}
\langle\langle-\theta_{q_1},\cdots,-\theta_{q_m}|\theta_{p_1},\cdots,\theta_{p_n}\rangle\rangle=f^{(o)}_n(\theta_{p_1},\cdots,\theta_{p_n})\delta_{m+n,N}
\delta_{\{q_1,\cdots,q_m\};\{p_1,\cdots,p_n\}},
\end{eqnarray} where $\delta_{\{q_1,\cdots,q_m\};\{p_1,\cdots,p_n\}}$ is defined as
\bea
\delta_{\{q_1,\cdots,q_m\};\{p_1,\cdots,p_n\}}=\left\{\begin{array}{ll}
1&{\rm if}\, \{ q_1,\cdots,q_m,p_1,\cdots,p_n\}=\{1,\cdots,N\},\\[6pt]
0&{\rm otherwise},\end{array}\rt.\label{Delta-XXX} \eea and
$f^{(o)}_n(\theta_{p_1},\cdots,\theta_{p_n})$ is given by
\begin{eqnarray}
f^{(o)}_n(\theta_{p_1},\cdots,\theta_{p_n})&=&\langle\langle-\theta_{p_{n+1}},\cdots,-\theta_{p_N}|\theta_{p_1},\cdots,\theta_{p_n}\rangle\rangle\no\\
&=&\prod_{j=1}^n(-1)^N\bar K_{21}^-(\theta_{p_j})d(-\theta_{p_j}-\eta)a(\theta_{p_j})\no\\
&&\times\prod_{k=n+1}^N(-1)^{N}
\bar K_{21}^-(-\theta_{p_k})a(-\theta_{p_k})d(\theta_{p_k}-\eta)\no\\
&&\times\prod_{j=1}^n\prod_{l>j}^n\frac{\theta_{p_j}+\theta_{p_l}}
{\theta_{p_j}+\theta_{p_l}+\eta}\,\prod_{j=n+1}^N
\prod_{l>j}^N\frac{\theta_{p_{j}}+\theta_{p_{l}}}{\theta_{p_{j}}+\theta_{p_{l}}-\eta}\no\\
&&\times\prod_{j=1}^n\prod_{l=n+1}^N\frac{\theta_{p_l}-\theta_{p_j}}{\theta_{p_l}-\theta_{p_j}-\eta}.
\label{f-function-XXX}
\end{eqnarray}
The right states
$\{|\theta_{p_1},\cdots,\theta_{p_n}\rangle\rangle\}$ given by
(\ref{Open-right-state}) (or the left states  $\{\langle\langle
-\theta_{p_1},\cdots,-\theta_{p_n}\}$ given by
(\ref{Open-left-state}))  form a right (or left) basis of the
Hilbert space. Therefore, any right (or left) state can be
decomposed as a unique linear combination of the basis.

%%%%%%%%%%%%%%%%%%%%%%%%%%%%%%%%%%%%%%%%%%%%%%%%%%%%%%%%%%%%%%%%%%%%%%%%%%%

\subsection{Retrieving the Bethe states}

Let $\langle\langle \Psi|$ be a common eigenstate of the transfer
matrix $t^{(o)}(u)$, namely, \bea \langle\langle
\Psi|\,t^{(o)}(u)=\langle\langle \Psi|\,\Lambda^{(o)}(u),\no \eea
where the eigenvalue $\Lambda^{(o)}(u)$ is given by (\ref{T-Q-XXX}).
Following the method used in \cite{Cao1}, we introduce \bea \bar
F_n(\theta_{p_1},\cdots,\theta_{p_n})=\langle\langle\Psi|\theta_{p_1},\cdots,\theta_{p_n}\rangle\rangle,\,n=0,\cdots,N,\,\quad
1\leq p_1<p_2<\cdots<p_n\leq N.\label{Components-1} \eea All these
quantities uniquely determine the eigenstate. Let us consider the
quantity of $\langle\langle\Psi|t(\theta_{p_{n+1}})
|\theta_{p_1},\cdots,\theta_{p_n}\rangle\rangle$. After a tedious
calculation, we obtain the following recursive  relations
\begin{eqnarray}
\Lambda^{(o)}(\theta_{p_{n+1}})\bar
F_{n}(\theta_{p_{1}},\cdots,\theta_{p_{n}})
\hspace{-0.08truecm}=\hspace{-0.08truecm}\frac{(2\theta_{p_{n+1}}\hspace{-0.18truecm}+\hspace{-0.08truecm}\eta
)\bar
K_{11}^+(\theta_{p_{n+1}})\hspace{-0.08truecm}+\hspace{-0.08truecm}\eta\bar
K_{22}^+(\theta_{p_{n+1}})} {2\theta_{p_{n+1}}+\eta}\bar
F_{n+1}(\theta_{p_{1}},\cdots,\theta_{p_{n+1}}).\label{Recursive-XXX}
\end{eqnarray} The above  relation allows us to determine $\{\bar F_n(\theta_{p_1},\cdots,\theta_{p_n})\}$ as
\bea \bar
F_n(\theta_{p_1},\cdots,\theta_{p_n})=\lt\{\prod_{j=1}^n\frac{(2\theta_{p_j}+\eta)\Lambda^{(o)}(\theta_{p_j})}
{(2\theta_{p_j}+\eta)\bar K^+_{11}(\theta_{p_j})+\eta\bar
K^+_{22}(\theta_{p_j})}\rt\}\,\bar F_0, \no \eea where $\bar
F_0=\langle\langle \Psi|\Omega\rangle_{\xi}$ is an overall scalar
factor. With (\ref{T-Q-XXX}) in mind, we further rewrite the above
expression as follows \bea
\bar F_n(\theta_{p_1},\cdots,\theta_{p_n})&=&\langle\langle\Psi|\theta_{p_1},\cdots,\theta_{p_n}\rangle\rangle\no\\[2pt]
&=&\lt\{\prod_{j=1}^n(-1)^N(\theta_{p_j}+p)\,a(\theta_{p_j})d(-\theta_{p_j}-\eta)
\frac{Q(\theta_{p_j}-\eta)}{Q(\theta_{p_j})}\rt\}\,\bar F_0,\no\\[2pt]
n&=&0,\cdots,N,\,\quad 1\leq p_1<p_2<\cdots<p_n\leq
N.\label{Components-2} \eea

The definitions (\ref{Dual-up})  and (\ref{Mon-XXX-1}) of the state $\langle 0|$ and the operators $A(u)$, $B(u)$, $C(u)$ and $D(u)$ allow us
to derive the following relations
\bea
 \langle 0|\,A(u)=a(u)\langle 0|,\quad \langle 0|\,D(u)=d(u)\langle 0|,\quad \langle 0|\,B(u)=0,\quad \langle 0|\,C(u)\neq 0,\label{Actions-1}
\eea where the functions $a(u)$ and $d(u)$ are given by
(\ref{a-b-functions-rational}). The expression (\ref{2-row-Mon-1})
of the double-row monodromy matrix  ${\mathbb{T}}(u)$ leads to the
actions (see (\ref{D-actions-1})-(\ref{D-actions-4}) ) of the matrix
elements of  ${\mathbb{T}}(u)$  on the state $ \langle 0| $. The
relations (\ref{2-row-Mon-2}) between the matrix elements of $\bar
{\mathbb{T}}(u)$  and those of ${\mathbb{T}}(u)$ allow us to derive
the following expressions of $\{\langle
0|\theta_{p_1},\cdots,\theta_{p_n}\rangle\rangle\}$ \bea \langle
0|\theta_{p_1},\cdots,\theta_{p_n}\rangle\rangle&=&
\lt\{\prod_{j=1}^n(-1)^N(\theta_{p_j}+p)\,a(\theta_{p_j})d(-\theta_{p_j}-\eta)\rt\}\,\langle 0|\Omega\rangle_{\xi},\no\\[4pt]
n&=&0,\cdots,N,\,\quad 1\leq p_1<p_2<\cdots<p_n\leq
N.\label{Components-3} \eea The proof of the above expression is
relegated to Appendix B. With the help of (\ref{State-1}) and
(\ref{Guage-Ref}), it is easy to check that for a generic nonzero
$\xi$ the overall constant $\langle 0|\Omega\rangle_{\xi}$ does not
vanish. For each solution of the BAEs (\ref{BAE-XXX}), let us
introduce the following left Bethe states \bea {}_B\langle
\l_1,\cdots,\l_N|=\langle 0|\, \lt\{\prod_{j=1}^N\frac{\bar
{\cal{C}}(\l_j)}{(-1)^N\bar
K^-_{21}(\l_j)d(\l_j)d(-\l_j-\eta)}\rt\}.\label{Bethe-state-open-1}
\eea The relations (\ref{C-eigenstate-1}), (\ref{C-eigenvalue-1})
and (\ref{Components-3}) imply that \bea {}_B\langle
\l_1,\cdots,\l_N|\theta_{p_1},\cdots,\theta_{p_n}\rangle\rangle&=&
\lt\{\prod_{j=1}^n(-1)^N(\theta_{p_j}+p)\,a(\theta_{p_j})d(-\theta_{p_j}-\eta)\frac{Q(\theta_{p_j}-\eta)}{Q(\theta_{p_j})}\rt\}\,
\langle 0|\Omega\rangle_{\xi},\no\\
n&=&0,\cdots,N,\,\quad 1\leq p_1<p_2<\cdots<p_n\leq
N.\label{Components-4} \eea Comparing the above expression with
(\ref{Components-2}), we conclude that the Bethe state ${}_B\langle
\l_1,\cdots,\l_N|$ given by (\ref{Bethe-state-open-1}) is an
eigenstate of the transfer matrix $t^{(o)}(u)$ with the
corresponding eigenvalue (\ref{T-Q-XXX}), provided that the
parameters $\{\l_j\}$ satisfy the BAEs (\ref{BAE-XXX}).

With the same procedure, we can construct the right Bethe state (up to an irrelevant scalar factor) as
\bea
 |\l_1,\cdots,\l_N\rangle_B=\prod_{j=1}^N\bar {\cal{B}}(\l_j) |0\rangle,\label{Bethe-state-open-2}
\eea
which is exactly the eigenstate conjectured in \cite{Bel13}.

It follows from their definitions that the two ``reference" states
$|0\rangle$ and $\langle 0|$ are independent of the inhomogeneous
parameters $\{\theta_j\}$ and therefore the Bethe state ${}_B\langle
\l_1,\cdots,\l_N|$ (or $|\bar \l_1,\cdots,\bar \l_N\rangle_B$) has a
well-defined homogeneous limit.

%%%%%%%%%%%%%%%%%%%%%%%%%%%%%%%%%%%%%%%%%%%%%%%%%%%%%%%%%%%%%%%
%                                                             %
%  7. Conclusions                                             %
%                                                             %
%                                                             %
%                                                             %
%%%%%%%%%%%%%%%%%%%%%%%%%%%%%%%%%%%%%%%%%%%%%%%%%%%%%%%%%%%%%%%

\section{Conclusions}
\label{Con} \setcounter{equation}{0}

In conclusion, a systematic method to retrieve the Bethe-type eigenstates of quantum integrable models is proposed. As examples, the eigenstates of
the XXZ spin-$\frac{1}{2}$ torus model and the open Heisenberg chain with generic boundary fields are derived based on the ODBA
solutions.

It should be remarked
that constructing the Bethe-type eigenstates of generic quantum integrable models without obvious reference state had challenged for many years. The present method provides an efficient way to retrieve reference states based on inhomogeneous
$T-Q$ relation that can be derived via ODBA. Naturally, this method can be generalized to other
integrable models without obvious reference state. For a given monodromy matrix, some mutually
commutative elements $T^{ij}(u)$, i.e., $[T^{ij}(u), T^{ij}(v)]$,
exist. The eigenstates of $T^{ij}(u)$ thus form an orthogonal and
complete basis (Sklyanin's SoV basis) of the Hilbert space. This basis together with the
$T-Q$ relation constructed from ODBA and the commutation relations
among the elements of the monodromy matrix, allow us to retrieve the Bethe-type
eigenstates of the transfer matrix step by step even without using a trial initial state.

%%%%%%%%%%%%%%%%%%%%%%%%%%%%%%%%%%%%%%%%%%%%%%%%%%%%%%%%%%%%%%%
%                                                             %
%  Acknowledgments                                            %
%                                                             %
%%%%%%%%%%%%%%%%%%%%%%%%%%%%%%%%%%%%%%%%%%%%%%%%%%%%%%%%%%%%%%%
\section*{Acknowledgments}

We would like  thank R.\,I. Nepomechie  for his valuable discussions.
This work was financial supported by NSFC under grant Nos. 11375141, 11374334, 11434013, 11425522, the 973 project under grant
No.2011CB921700, BCMIIS and the Strategic Priority Research Program
of CAS.

%%%%%%%%%%%%%%%%%%%%%%%%%%%%%%%%%%%%%%%%%%%%%%%%%%%%%%%%%%%%%%%%
%                                                             %
%  Appendix A                                                 %
%                                                             %
%                                                             %
%                                                             %
%%%%%%%%%%%%%%%%%%%%%%%%%%%%%%%%%%%%%%%%%%%%%%%%%%%%%%%%%%%%%%%

\section*{Appendix A: Proof of (\ref{reference-state}) satisfying (\ref{refernece-conditions}) }
\setcounter{equation}{0}
\renewcommand{\theequation}{A.\arabic{equation}}

Let us introduce the following inner product
\begin{eqnarray}
\langle\theta_{1},\cdots,\theta_{n}|\prod_{k=1}^m\bar B(u_k)|0\rangle=\delta_{n,m}\,g_n(\{\theta_j\}|\{u_{\alpha}\}),
\quad g_0=\langle0|0\rangle=1.\label{4-g-functions-1}
\end{eqnarray}
The relations (\ref{RLL-4})-(\ref{RLL-6}) allow us to derive some
recursive relations for the function $g_n(\{\theta_j\} |$ $
\{u_{\alpha}\})$:
\begin{eqnarray}
g_n(\{\theta_j\}|\{u_{\alpha}\})&=&\sum_{l=1}^n\frac{\sinh\eta\,\bar d(u_1) \bar a(\theta_l)}{\sinh(u_1-\theta_l)}
\lt\{\prod_{j\neq l}^n\frac{\sinh(u_1-\theta_j+\eta)}{\sinh(u_1-\theta_j)}\rt.\no\\
&& \times
\lt.\frac{\sinh(\theta_l-\theta_j-\eta)}{\sinh(\theta_l-\theta_j)}\rt\}
g_{n-1}(\{\theta_j\}_{j\neq l}|\{u_{\alpha}\}_{\alpha\neq
1}).\label{4-recursive-relations}
\end{eqnarray}
Moreover, let us introduce the following quantity
\bea
\bar g_n(\{\theta_{p_j}\})=[n]_q!\,\langle\theta_{p_1},\cdots,\theta_{p_n}|\Omega;\{\theta_j\}\rangle
=\langle\theta_{p_1},\cdots,\theta_{p_n}|\,\lt(\tilde{B}^-\rt)^n|0\rangle.\label{4-product}
\eea The definitions (\ref{4-g-functions-1}) and (\ref{4-limit}) imply that one can calculate the function
$\bar g_n(\{\theta_{j}\})$ by the following limit
\bea
\hspace{-0.8truecm}\bar g_n(\{\theta_{j}\})=\lim_{\{u_l\to +\infty\}} \lt\{
\lt[\prod_{l=1}^n\lt(2\sinh\eta\,e^{-u_l}\rt)^{N-1}e^{\sum_{k=1}^N\theta_k}\rt]
g_n(\{\theta_j\}|\{u_{l}\})\rt\}.
\eea
Keeping the recursive relations (\ref{4-recursive-relations}) in mind, we can derive the following recursive relations
\bea
\bar g_n(\{\theta_{j}\})&=&\sum_{l=1}^ne^{(n-1)\eta}\,a(\theta_l)e^{\theta_l}\,
\prod_{j\neq l}^n\frac{\sinh(\theta_l-\theta_j-\eta)}{\sinh(\theta_l-\theta_j)}\,
\bar g_{n-1}(\{\theta_{j}\}_{j\neq l}),\no\\
n&=&1,\cdots,N,\no
\eea with the initial condition of $\bar g_0=1$. The above recursive relations uniquely determine the functions
$\{\bar g_n(\{\theta_{j}\})|n=0,\cdots,N\}$:
\bea
\bar g_n(\{\theta_{j}\})=\lt\{\prod_{l=1}^n\,a(\theta_l)e^{\theta_l}\rt\}\,[n]_q!,\quad n=0,\cdots,N. \label{4-g-functions-4}
\eea Note that the following identities were used in deriving the above equations
\bea
\sum_{l=1}^ne^{(n-1)\eta}\,\prod_{j\neq l}^n\frac{\sinh(\theta_l-\theta_j-\eta)}{\sinh(\theta_l-\theta_j)}
=1+e^{2\eta}+\cdots+e^{2(n-1)\eta}=[n]_q.
\eea

Substituting (\ref{4-g-functions-4}) into
(\ref{4-product}), we find that the state $|\Omega;\{\theta_j\}\rangle$ given by (\ref{reference-state}) indeed satisfies the relations
(\ref{refernece-conditions}).

%%%%%%%%%%%%%%%%%%%%%%%%%%%%%%%%%%%%%%%%%%%%%%%%%%%%%%%%%%%%%%%%
%                                                             %
%  Appendix B                                                 %
%                                                             %
%                                                             %
%                                                             %
%%%%%%%%%%%%%%%%%%%%%%%%%%%%%%%%%%%%%%%%%%%%%%%%%%%%%%%%%%%%%%%

\section*{Appendix B: Proof of  (\ref{Components-3}) }
\setcounter{equation}{0}
\renewcommand{\theequation}{B.\arabic{equation}}

The gauged double-row monodromy matrix $\bar {\mathbb{T}}(u)$ given
by (\ref{2-row-Mon-2}) satisfies the reflection algebra \cite{Skl88}
\begin{eqnarray}
R_{12}(u-v)\bar{\mathbb{T}}_1(u)R_{21}(u+v)\bar{\mathbb{T}}_2(v)
=\bar{\mathbb{T}}_2(v)R_{12}(u+v)\bar{\mathbb{T}}_1(u)R_{21}(u-v),
%\label{Reflection-alg}
\end{eqnarray} which leads to the following  relevant relations
\begin{eqnarray}
\bar{\cal{C}}(u)\bar{\cal{A}}(v)&=&\frac{(u+v)(u-v+\eta)}{(u-v)(u+v+\eta)}\bar{\cal{A}}(v)\bar{\cal{C}}(u)
                  -\frac{\eta}{u+v+\eta}\bar{\cal{D}}(u)\bar{\cal{C}}(v)\no\\[2pt]
   && -\frac{(u+v)\eta}{(u-v)(u+v+\eta)}\bar{\cal{A}}(u)\bar{\cal{C}}(v),\label{XXX-RLL-6}\\[2pt]
\bar{\cal{D}}(v)\bar{\cal{C}}(u)&=&\frac{(u+v)(u-v+\eta)}{(u-v)(u+v+\eta)}\bar{\cal{C}}(u)\bar{\cal{D}}(v)
                  -\frac{\eta}{u+v+\eta}\bar{\cal{C}}(v)\bar{\cal{A}}(u)\no\\[2pt]
   &&  -\frac{(u+v)\eta}{(u-v)(u+v+\eta)}\bar{\cal{C}}(v)\bar{\cal{D}}(u),\label{XXX-RLL-7}\\[2pt]
\bar{\cal{A}}(u)\bar{\cal{A}}(v)&=&\bar{\cal{A}}(v)\bar{\cal{A}}(u)+\frac{\eta}{u+v+\eta}
                  \bar{\cal{B}}(v)\bar{\cal{C}}(u)-\frac{\eta}{u+v+\eta}\bar{\cal{B}}(u)\bar{\cal{C}}(v),\label{XXX-RLL-8}\\[2pt]
\bar{\cal{D}}(u)\bar{\cal{D}}(v)&=&\bar{\cal{D}}(v)\bar{\cal{D}}(u)
                  +\frac{\eta}{u+v+\eta}\bar{\cal{C}}(v)\bar{\cal{B}}(u)
                  -\frac{\eta}{u+v+\eta}\bar{\cal{C}}(u)\bar{\cal{B}}(v),\label{XXX-RLL-9}\\[2pt]
\bar{\cal{D}}(u)\bar{\cal{A}}(v)&=&\bar{\cal{A}}(v)\bar{\cal{D}}(u)
                  -\frac{(u+v+2\eta)\eta}{(u-v)(u+v+\eta)}\bar{\cal{B}}(u)\bar{\cal{C}}(v)\no\\[2pt]
   &&+\frac{(u+v+2\eta)\eta}{(u-v)(u+v+\eta)}\bar{\cal{B}}(v)\bar{\cal{C}}(u).\label{XXX-RLL-10}
\end{eqnarray}
Equations (\ref{U-matrix}) and (\ref{2-row-Mon-2}) imply that the following relations hold:
\bea
 \bar {\cal{A}}(u)&=&\frac{1}{2\xi\sqrt{1+\xi^2}}\lt\{\xi (1+\sqrt{1+\xi^2}){\cal{A}}(u)+\xi^2{\cal{C}}(u)\rt.\no\\[2pt]
       && +\lt.\xi^2{\cal{B}}(u)-\xi (1-\sqrt{1+\xi^2}){\cal{D}}(u) \rt\},\label{D-relation-1}\\[2pt]
\bar {\cal{C}}(u)&=&\frac{1}{2\xi\sqrt{1+\xi^2}}\lt\{\xi (1+\sqrt{1+\xi^2}){\cal{A}}(u)-(1+\sqrt{1+\xi^2})^2{\cal{C}}(u)\rt.\no\\[2pt]
       && +\lt.\xi^2{\cal{B}}(u)-\xi (1+\sqrt{1+\xi^2}){\cal{D}}(u) \rt\},\label{D-relation-2}\\[2pt]
\bar {\cal{D}}(u)&=&\frac{1}{2\xi\sqrt{1+\xi^2}}\lt\{\xi (\sqrt{1+\xi^2}-1){\cal{A}}(u)-\xi^2{\cal{C}}(u)\rt.\no\\[2pt]
       &&  -\lt.\xi^2{\cal{B}}(u)+\xi (1+\sqrt{1+\xi^2}){\cal{D}}(u) \rt\}.
\eea We note that \bea
 \langle 0|\,{\cal{A}}(u)&=&(-1)^NK^-_{11}(u)\,a(u)\,d(-u-\eta)\,\langle 0|,\label{D-actions-1}\\[2pt]
 \langle 0|\,{\cal{D}}(u)&=&(-1)^N\frac{\eta }{2u+\eta}K^-_{11}(u)\,a(u)\,d(-u-\eta)\,\langle 0|\no\\[2pt]
         &&  +(-1)^N\frac{(2u+\eta)K^-_{22}(u)-\eta K^-_{11}(u)}{2u+\eta}\,d(u)\,a(-u-\eta)\,\langle 0|,\label{D-actions-2}\\[2pt]
 \langle 0|\,{\cal{B}}(u)&=&0,\label{D-actions-3}\\[2pt]
 \langle 0|\,{\cal{C}}(u)&=&(-1)^N\frac{2u}{2u+\eta}K^-_{11}(u)\,d(-u-\eta)\,\langle 0|\,C(u)\no\\[2pt]
         &&  +(-1)^N\frac{\eta K^-_{11}(u)-(2u+\eta)K^-_{22}(u)}{2u+\eta}\,d(u)\,\langle 0|\,C(-u-\eta).\label{D-actions-4}
\eea

Let us consider the quantity $\langle 0|\,\bar
{\cal{C}}(\theta_{p_{n+1}})\,|\theta_{p_1},\cdots,\theta_{p_n}\rangle\rangle$.
Acting $\bar {\cal{C}}(\theta_{p_{n+1}})$ to the right, from the
relations (\ref{C-eigenstate-1}) and (\ref{C-eigenvalue-1}), we know
that $\langle 0|\,\bar
{\cal{C}}(\theta_{p_{n+1}})\,|\theta_{p_1},\cdots,\theta_{p_n}\rangle\rangle$
vanishes. On the other hand, acting it to the left and using
(\ref{D-actions-1})-(\ref{D-actions-4}) and the relation
(\ref{D-relation-2}), we  obtain \bea
 \langle 0|\,C(\theta_{p_{n+1}})\,|\theta_{p_1},\cdots,\theta_{p_n}\rangle\rangle=\frac{\xi}{1+\sqrt{1+\xi^2}}\,a(\theta_{p_{n+1}})\,
      \langle 0|\theta_{p_1},\cdots,\theta_{p_n}\rangle\rangle,\quad j=1,\cdots,N.\label{D-relations-4}
\eea From the definition (\ref{Open-right-state}) we have \bea
\langle
0|\theta_{p_1},\cdots,\theta_{p_{n+1}}\rangle\rangle&=&\langle
0|\,\bar {\cal{A}}(\theta_{p_{n+1}})\,
         |\theta_{p_1},\cdots,\theta_{p_{n}}\rangle\rangle.\no
\eea  Acting $\bar {\cal{A}}(\theta_{p_{n+1}})$ to the left and
using (\ref{D-actions-1})-(\ref{D-actions-4}) and the relation
(\ref{D-relation-1}), we have \bea \langle
0|\theta_{p_1},\cdots,\theta_{p_{n+1}}\rangle\rangle\hspace{-0.28truecm}&=&\hspace{-0.28truecm}
          \frac{(-1)^N}{2\sqrt{1\hspace{-0.08truecm}+\hspace{-0.08truecm}\xi^2}}\lt\{
          (1\hspace{-0.12truecm}+\hspace{-0.12truecm}\sqrt{1\hspace{-0.08truecm}+\hspace{-0.08truecm}\xi^2})
          K^-_{11}(\theta_{p_{n+1}})a(\theta_{p_{n+1}})d(\hspace{-0.04truecm}-\hspace{-0.04truecm}\theta_{p_{n+1}}
          \hspace{-0.08truecm}-\hspace{-0.08truecm}\eta)\langle 0
          |\theta_{p_1},\cdots,\theta_{p_{n}}\rangle\rangle\rt.\no\\[4pt]
   && -(1\hspace{-0.12truecm}-\hspace{-0.12truecm}\sqrt{1\hspace{-0.08truecm}+\hspace{-0.08truecm}\xi^2})
          \frac{\eta}{2\theta_{p_{n+1}}\hspace{-0.08truecm}+\hspace{-0.08truecm}\eta}K^-_{11}(\theta_{p_{n+1}})a(\theta_{p_{n+1}})
          d(\hspace{-0.04truecm}-\hspace{-0.04truecm}\theta_{p_{n+1}}
          \hspace{-0.08truecm}-\hspace{-0.08truecm}\eta)\langle 0
          |\theta_{p_1},\cdots,\theta_{p_{n}}\rangle\rangle\no\\[4pt]
   && +\xi\frac{2\theta_{p_{n+1}}}{2\theta_{p_{n+1}}+\eta}
          \lt.K^-_{11}(\theta_{p_{n+1}})d(\hspace{-0.04truecm}-\hspace{-0.04truecm}\theta_{p_{n+1}}
          \hspace{-0.08truecm}-\hspace{-0.08truecm}\eta)\langle 0
          |C(\theta_{p_{n+1}})|\theta_{p_1},\cdots,\theta_{p_{n}}\rangle\rangle\rt\}\no\\[4pt]
&=&(-1)^NK^-_{11}(\theta_{p_{n+1}})a(\theta_{p_{n+1}})
         d(-\theta_{p_{n+1}}-\eta)\langle 0|\theta_{p_1},\cdots,\theta_{p_{n}}\rangle\rangle,\no\\[4pt]
  n&=& 0,\cdots, N-1.\label{D-relation-6}
\eea

%%%%%%%%%%%%%%%%%%%%%%%%%%%%%%%%%%%%%%%%%%%%%%%%%%%%%%%%%%%%%%%
%                                                             %
%  References                                                 %
%                                                             %
%%%%%%%%%%%%%%%%%%%%%%%%%%%%%%%%%%%%%%%%%%%%%%%%%%%%%%%%%%%%%%%

\end{document}